\documentclass{pramana}

\usepackage{graphicx,amsmath,bm}
\usepackage{color}
\usepackage{hyperref}

\begin{document}

\title{Interplay of intra- and inter-dependence affects the robustness of network
of networks}


\author{Aradhana Singh\textsuperscript{1,*} \and Sitabhra Sinha\textsuperscript{1,2}}
\affilOne{\textsuperscript{1} The Institute of Mathematical Sciences, CIT Campus, Taramani, Chennai 600113, India\\}
\affilTwo{\textsuperscript{2} Homi Bhabha National Institute, Anushaktinagar, Mumbai 400094, India}


\twocolumn[{

\maketitle

\corres{aradhana22singh@gmail.com}


\begin{abstract} The existence of inter-dependence between multiple networks imparts an additional scale of complexity to such systems often referred to as ``network of networks'' (NON). We have investigated the robustness of NONs to random breakdown
of their components, as well as targeted attacks, as a function of the relative
proportion of intra- and inter-dependence among the constituent networks.
We focus on bi-layer networks with the two layers comprising different number of nodes
in general and where the ratio of intra-layer to inter-layer connections, $r$,
can be varied, keeping the total number of nodes and overall connection density invariant. We observe that while the responses of the different networks to random breakdown of nodes are similar, dominantly intra-dependent networks ($r\ll1$)
are robust with respect to attacks that target nodes having highest degree but when
nodes are removed on the basis of highest betweenness centrality (CB), they exhibit
a sharp decrease in the size of the largest connected component (resembling a first
order phase transition) followed by a more gradual decrease as more nodes are removed
(akin to a second order transition).
As $r$ is increased resulting in the network becoming strongly inter-dependent ($r\gg1$), we observe that this hybrid nature of the transition in the size of the largest 
connected component
in response to targeted node removal (based on highest CB) changes to a purely
continuous or second-order transition.  
We also explore the role of layer size heterogeneity on robustness, finding that
for a given $r$ having layers comprising very different number of nodes results
in a bimodal degree distribution. For dominantly inter-dependent networks, this
results in the nodes of the smaller layer becoming structurally central. 
Selective removal of these nodes, which constitute a relatively small 
fraction of the network, leads to breakdown of the entire system - making the
inter-dependent networks even more fragile to targeted attacks than 
scale-free networks having power-law degree distribution.  
\end{abstract}



}]



\section{Introduction}
Robustness, a property often attributed to complex systems occurring in nature,
refers to their ability to maintain most of their vital functions even
when subjected to noise or perturbations, both extrinsic and intrinsic,
that may result in loss or damage of a significant fraction of their 
components~\cite{Erica_Jen}. 
The investigation of robust systems, especially those that occur in
biology and ecology, with the aim of identifying the features that contribute
to their ability to withstand component failures or attacks on parts 
thereof, have obvious implications in terms of applications.
These include designing robust man-made systems, as well as, arriving at
fail-safe strategies to reduce vulnerabilities of existing systems
such as the electrical power grid, where an initially
small local perturbation (such as shorting caused by a branch falling on 
a transmission line) can occasionally trigger a massive system-wide
breakdown resulting in power blackouts over entire regions~\cite{Powergrid}.
As many complex systems can be represented as networks, with the
components represented as nodes while the interactions between them
are represented as links, robustness can also be measured in terms of
the ability of a system to maintain its integrity even after
a specified fraction of its nodes and/or links have been 
removed~\cite{randomnet_percolation,Barabasi_robustness,Barabasi_book}.  
An oft-cited example is the internet, comprising servers (nodes)
connected by data cables (links), whose functioning
should not be affected significantly by temporary
loss of components through failures occurring randomly, as well
as, malicious denial-of-service attacks that may target specific 
nodes~\cite{Internet_robustness_pnas}. Following the 2007-9 financial crisis, the robustness
of the network of financial institutions has also been the
subject of intense investigation by scientists who seek to
understand factors contributing to systemic risk that
can cause credit default by a few firms to eventually result
in an overall economic catastrophe~\cite{May_Ecology_for_Bankers}.

Complicating the already difficult question of what factors lead to
robustness of complex networks is the fact that in reality, most networks
do not operate completely in isolation but often are seen to interact 
with other equally complex networks.
Moreover, inter-dependence between multiple networks could be a crucial
feature underlying the proper functioning of each of them. 
An example is the coupled system of the electrical power grid and the
communication network of computers~\cite{Havlin_nature2010}. 
While the network of computers control the functioning of the power grid, the
computers are dependent on the grid for their power. Failure in nodes
of one of the networks (e.g., shutting down of a power generation unit) 
would affect nodes in the other network (e.g., disrupting the 
communication between computers), which in turn will lead to further
breakdowns of both the networks in a recursive 
fashion~\cite{Havlin_book,Havlin_nature2011,Parera}.
In general, inter-dependent networks can be seen as comprising different
layers in a composite network of networks (NON). 

A strikingly novel aspect of inter-dependent networks is that they typically
respond very differently to structural perturbations such as removal
of a fraction of their nodes when compared to the behavior of the component 
networks in isolation. 
In particular, inter-dependent networks exhibit a first order
phase transition in the size of the largest connected component when
nodes are gradually removed, which changes to a continuous transition
when the fraction of inter-dependent nodes is reduced~\cite{Havlin_prl2010}.
Assuming that only nodes belonging to the largest connected component
remain functional, this would suggest that inter-dependent networks are
more vulnerable to node failure and targeted attacks than the
individual systems that they
comprise~\cite{Havlin_nature2010, inter-dependent_attack}.
While a few earlier studies have considered the role of intra-, as well as,
inter-network dependences in determining the robustness of 
NONs~\cite{Havlin_pre2011,Dsouza,Singh2017}, it is important to
keep the average degree of the nodes invariant when comparing 
systems with different ratios of intra- to inter-network connections
(as otherwise we cannot disambiguate the contribution of the overall
number of connections from that specifically of the inter-dependent 
links). In addition, the different networks have often been chosen
to be of the same size. However, in reality, NONs can comprise 
component networks comprising widely differing number of nodes.
In this paper we report the results of a systematic investigation of
the robustness of NONs to different types of node removal strategies,
incorporating the different aspects mentioned above.

\begin{figure}[h!]
\includegraphics[width=0.99\linewidth,clip]{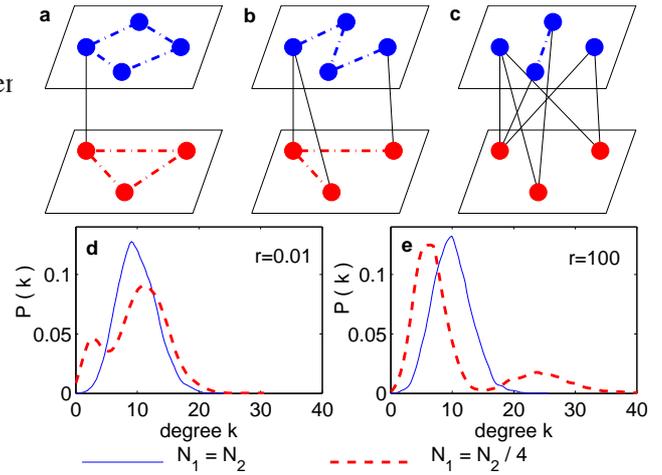}
\caption{(Color online) {\bf Network of networks consisting of two layers of 
nodes that have connections both within a layer (broken lines) and
between layers (solid lines) representing intra- and inter-dependence,
respectively, exhibit very different degree distributions depending
on relative sizes of the layers.}
(a-c) Schematic diagram of bi-layer networks corresponding to situations
where the networks are dominantly
intra-dependent (a), homogeneous (b) and dominantly inter-dependent. 
(d-e) Degree distributions for bi-layer networks having different
ratios of intra-layer to inter-layer connections, $r$ [viz., $r=0.01$ in (d)
and $r=100$ in (e)], 
where the
layers could be either of equal size, i.e., $N_1 = N_2$ (solid curves) or have 
unequal sizes, viz., $N_1 = N_{2}/4$ (broken curves). While the former case
does not show much variation between dominantly intra-dependent (i.e., $r \ll 1$)
and dominantly inter-dependent (i.e., $r \gg 1$) systems, NONs
with heterogeneous layer sizes show bimodal degree distributions
whose profiles differ for dominantly intra-dependent
and dominantly inter-dependent systems.
Each distribution is averaged over 10 realization with 
total network size $N = N_1 + N_2 = 500$ and average degree $\langle k \rangle = 10$.}
\label{schematic}
\end{figure}
\section{Model}
The model system we consider for our investigation is a NON of two networks
comprising $N_1$ and $N_2$ nodes, respectively. In order to analyze
the relative contributions of intra- and inter-dependence in this system,
we alter the probabilities of a connection between nodes belonging to the 
same layer ($p_{intra}$) and those belonging to different layers ($p_{inter}$).
This is done by assigning different values to the ratio $r = \frac{p_{inter}}{p_{intra}}$ 
while keeping the total size of the
NON ($N=N_1+N_2$) and the average degree of the network $\langle k \rangle$ 
invariant~\cite{Pan_2009}.
For $r \ll 1$, the NON is dominantly intra-dependent [Fig.~\ref{schematic}~(a)], while it is 
dominantly inter-dependent if $r \gg 1$ [Fig.~\ref{schematic}~(c)]. The special case of $r=1$ corresponds
to a homogeneous Erd\"os-Renyi network [Fig.~\ref{schematic}~(b)].
Thus, as $r$ is increased from $0$, the NON changes gradually from being 
completely intra-dependent (consisting of two isolated modules) in one limit to 
completely inter-dependent (corresponding to a bipartite network, which
can be viewed as a hierarchical network consisting of two levels) in the other
limit. 
  
Randomly connected bi-layer networks where the two layers are of the same
size ($N_1=N_2$) have Poisson degree distributions regardless of $r$ 
[Fig.~\ref{schematic}~(d-e), solid curves]. 
However, if $N_1$ and $N_2$ are very different,
this results in the two layers having very different average degrees
(even though average degree of the NON, $\langle k \rangle$, remains
unchanged) with the overall degree distribution exhibiting a bimodal form
[broken curves in Fig.~\ref{schematic}~(d-e)].
The exact profile of the bimodal distribution depends on the value of $r$,
with the lower peak corresponding to the smaller (larger) layer for
dominantly intra-dependent (inter-dependent) networks.

We have considered the robustness of the model bi-layer networks described
above using a standard percolation-theoretic approach~\cite{randomnet_percolation}. Specifically, we remove nodes one at a time using different
strategies, e.g., at random or choosing nodes having the highest degree
or betweenness centrality (CB). After removing a fraction $f$ of the $N$ nodes
in the NON, we measure the probability that a randomly chosen node is still 
part of the largest connected component (LCC) of the NON after these
removals [$P_{LCC} (f)$], by expressing it in 
terms of the probability that the node was part of the LCC of the NON before
any nodes were removed [$P_{LCC} (0)$].  
Note that, for a homogeneous Erd\"os-Renyi random network (for $r=1$),  it is well-known that even after removal of a fraction $f$ of 
the nodes 
[such that the effective size of the network is now $N_{eff} = (1-f)N$],
the Poisson character of the degree distribution is preserved with only
the effective average degree reducing to $k_{eff} = (1-f)k$.
As the condition for a Erd\"os-Renyi network to possess a giant component
is $\langle k^2 \rangle/\langle k \rangle \gtrsim 2$~\cite{Molloy-reed},
the critical value of fraction of nodes removed beyond which the network 
exhibits a transition to isolated fragments is given by $f_c = 1-(1/\langle k \rangle)$. This provides a natural benchmark against which to compare the
robustness of the random bi-layer networks in response to 
removal of
a fraction of their nodes. We have also compared the results with that
of the Price-Barabasi-Albert scale-free network that has been shown to
be more robust with respect to random removal of nodes compared to 
Erd\"os-Renyi networks, but extremely vulnerable to attacks targeted
at nodes having highest degree or CB~\cite{Barabasi_book}.
\begin{figure}[h!]
\begin{center}
\includegraphics[width=0.99\linewidth,clip]{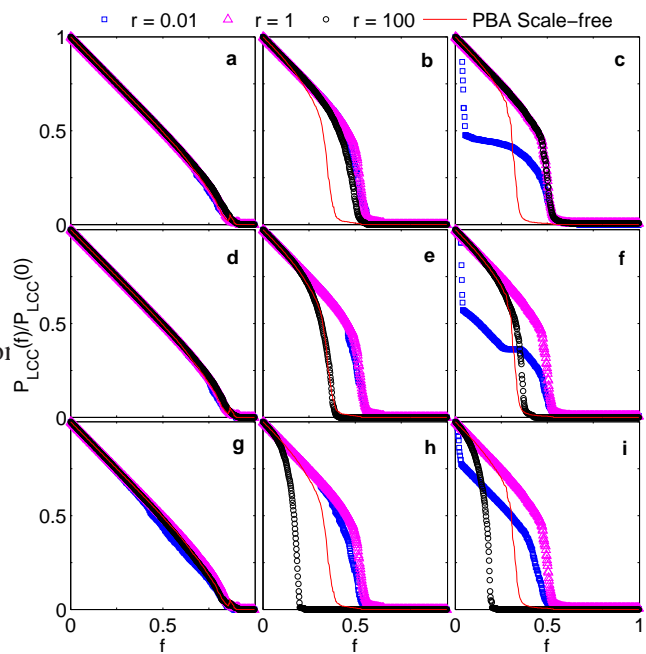}
\end{center}
\caption{(Color online) {\bf Robustness of random networks comprising two layers (with $N_1$ and $N_2$
nodes, respectively) that have different 
proportions of intra- and inter-layer dependence shown for different
types of node failures and layer size heterogeneity.}
On removing a fraction $f$ of the nodes in the network, the probability
$P_{LCC}(f)$ that
a node will be part of the largest connected component (LCC) is expressed
relative to the probability $P_{LCC}(0)$ that it was part of the LCC in 
the original network. The first row (panels a-c) shows the situation where
the two layers possess an identical number of nodes (i.e., $N_1 = N_2$), while
the second row (panels d-f) and third row (panels g-i) considers layers of
unequal size, viz., $N_1 = 2 N_2 / 3$ and  $N_1 = N_2 / 4$, respectively.
In all cases the total size of the network $N = N_1+N_2 = 500$ and average degree 
$\langle k \rangle = 10$. In each row, different panels show the robustness
of a network against different types of node failure protocols, corresponding
to removal of nodes at random (left), according to highest degree (center)
and according to highest betweenness centrality (right). 
Each panel shows the response to successive removal of nodes for networks
that are (i) dominantly intra-dependent ($r=0.01$, blue squares), (ii) 
homogeneous ($r=1$, maroon triangles) and (iii) dominantly inter-dependent
($r=100$, black circles). For comparison, we also show the response of
Price-Barabasi-Albert scale-free networks (red curve).
Each data point is obtained by averaging over $10$ network realizations.
We note that while for random breakdown of nodes, the response of the different
networks have similar profiles, with respect to attacks that target nodes
having highest degree or betweenness centrality, the dominantly inter-dependent networks are relatively
more vulnerable than the other types of networks
when the layers have very different sizes. 
In addition, the dominantly intra-dependent networks show a sharp decrease
in the fraction remaining in LCC for low $f$ when the attacks target nodes having highest betweenness centrality.
}
\label{robust_s0}
\end{figure}

\section{Results}
We first consider the response of bi-layer networks to removal of nodes chosen
at random for NONs characterized by different ratios of intra- and 
inter-dependence and where the layers are of same size 
[Fig.~\ref{robust_s0}~(a)].
We observe that regardless of $r$, the networks exhibit a similar response
profile to removal of nodes. A second-order transition is seen to occur at a 
critical value $f_c \sim 0.9$ of the fraction of nodes removed, where the
system reduces to several disconnected fragments. 
Introducing layer size heterogeneity does not appreciably alter the results
as can be seen from  panels (d) and (g) of Fig.~\ref{robust_s0} that correspond
to $N_1 = 2 N_2/3$ and $N_1 = N_2 /4$, respectively.
 
We next consider robustness of the NON against targeted attacks aimed
at structurally important nodes. These could either be the hubs, i.e.,
nodes having the highest degree, or may be connecting a large number of
nodes to each other through shortest paths that pass through them, i.e.,
nodes with highest CB~\cite{Barabasi_robustness}. 
We observe that dominantly intra-dependent networks are almost as robust as
Erd\"os-Renyi networks against attacks targeted at highest degree nodes,
while the dominantly inter-dependent networks are only marginally less
robust [Fig.~\ref{robust_s0}~(b)]. We observe that at around $f_c \sim 0.5$,
the networks exhibit a smooth transition to fragmentation. Note that,
the Price-Barabasi-Albert scale-free network is much less robust against
degree-based attacks and collapses at $f_c \sim 0.4$.
With increasing layer size 
heterogeneity however, 
the dominantly inter-dependent networks become increasingly 
fragile with the transition to fragmented state occurring at critical
values of $f$ that may be even lower than that for scale-free networks
[see panels (e) and (h) of Fig.~\ref{robust_s0}]. By contrast, intra-dependent
networks do not show any variation with respect to changing sizes 
of the layers.

Dominantly inter-dependent networks show a similar behavior when instead
of targeting highest degree nodes, highest BC nodes are removed preferentially
[panels (c), (f) and (i) of Fig.~\ref{robust_s0}]. However, 
the dominantly intra-dependent networks exhibit a strikingly different
response, with the size of the LCC showing a very
sharp decrease (resembling a first-order phase transition) from $N$
to $N_1$ upon removing only about $3\%$ of the nodes.
This suggests that at this value of $f$ ($\sim 0.03$), the layers of the
NON become isolated from each other. Following this, the effect of removing additional nodes according to highest CB is similar 
to that for Erd\"os-Renyi networks and consequently, we observe a continuous
transition to the fragmented state, explaining the hybrid phase transition
seen for the case of dominantly intra-dependent networks. 
Increasing layer size heterogeneity only changes this picture by decreasing
the critical value of $f$ at which the initial sharp decrease in the
LCC size occurs, as well as, the magnitude of the decrease.

The response of the dominantly inter-dependent networks with respect 
to targeted attacks on nodes (based either on highest degree or highest CB)
as layer size heterogeneity increases can be understood in terms of the 
changing connectivity profile as revealed by the degree distribution
[Fig.~\ref{schematic}~(e)]. When the two layers are similar in terms
of size, almost all nodes are equivalent in terms of their degree.
Thus, the response of the network to attacks will be almost
identical to that seen for Erd\"os-Renyi networks. However,
when the sizes of the two layers are
very different, the nodes of the smaller layer typically would have
much higher degree than the average degree of the NON [as
revealed by the bimodal degree distribution shown in panel (e) of Fig.~\ref{schematic}]. Thus, these will
function as hubs of the network. Targeting these relatively fewer number
of nodes will severely damage the network in terms of connectivity.
However, identifying such nodes in dominantly inter-dependent NONs and providing
them additional protection will be an efficient procedure for
increasing the robustness of the entire system.

\section{Conclusion}
In this paper we have reported the results of our investigation
on the role played by intra- and inter-dependence in imparting robustness 
to NONs by considering an ensemble of model random bi-layer networks. 
By systematically varying the relative density of intra- and inter-layer
connections we show that 
increasing inter-dependence can make such NONs vulnerable to targeted
attacks on nodes, especially when different layers are populated by
very different numbers
of nodes. This can be related to the very different connectivity profiles
of the nodes in the two layers, manifested in a bimodal degree distribution
for the NON.
We also observe that when faced with attacks targeted
at nodes having highest CB, increased dominant intra-dependence results in a
hybrid transition. This corresponds to an initially sharp decrease in the size
of the LCC (resembling a first-order phase transition) followed by a 
continuous or second-order transition with increasing fraction of nodes
removed. 
As in NONs occurring in nature the sizes of the different
component networks can be quite different, our results may provide 
insights into their robustness and help in suggesting guidelines for 
constructing more robust artificial NONs.  

\section*{Acknowledgments}
We would like to thank Shakti N. Menon for helpful discussions.
This work was supported in part by IMSc Complex Systems
(XII Plan) Project funded by the Department of
Atomic Energy, Government of India.
We thank the IMSc
High Performance Computing facility for access to the Nandadevi
cluster in which the simulations required for this work were done.

\end{document}